\documentclass[twocolumn]{aastex631}
\usepackage{color}
\usepackage{graphicx}
\usepackage{amsmath}
\usepackage{multirow}
\usepackage{threeparttable}
\usepackage{bm}
\usepackage[section]{placeins}

\usepackage{makecell} 

\begin{document}
\title{New tests of cosmic distance duality relation with DESI 2024 BAO observations}
\author{Qiumin Wang}
\affiliation{Institute for Frontiers in Astronomy and Astrophysics, Beijing Normal University, Beijing 102206, China;}
\affiliation{School of Physics and Astronomy, Beijing Normal University, Beijing 100875, China;}

\author{Shuo Cao$^{\ast}$}
\affiliation{Institute for Frontiers in Astronomy and Astrophysics, Beijing Normal University, Beijing 102206, China;}
\affiliation{School of Physics and Astronomy, Beijing Normal University, Beijing 100875, China;}
\email{$\ast$ caoshuo@bnu.edu.cn}

\author{Jianyong Jiang}
\affiliation{School of Physics and Astronomy, Beijing Normal University, Beijing 100875, China;}

\author{Kaituo Zhang}
\affiliation{Department of Physics, Anhui Normal University, Wuhu, Anhui 241000, China}

\author{Xinyue Jiang}
\affiliation{Institute for Frontiers in Astronomy and Astrophysics, Beijing Normal University, Beijing 102206, China;}
\affiliation{School of Physics and Astronomy, Beijing Normal University, Beijing 100875, China;}

\author{Tonghua Liu$\dagger$}
\affiliation{School of Physics and Optoelectronic, Yangtze University, Jingzhou 434023, China}
\email{$\dagger$ liutongh@yangtzeu.edu.cn}

\author{Chengsheng Mu}
\affiliation{Institute for Frontiers in Astronomy and Astrophysics, Beijing Normal University, Beijing 102206, China;}
\affiliation{School of Physics and Astronomy, Beijing Normal University, Beijing 100875, China;}

\author{Dadian Cheng}
\affiliation{Institute for Frontiers in Astronomy and Astrophysics, Beijing Normal University, Beijing 102206, China;}
\affiliation{School of Physics and Astronomy, Beijing Normal University, Beijing 100875, China;}

\begin{abstract}
In this paper, we test the cosmic distance duality relation (CDDR), as required by the Etherington reciprocity theorem, which connects the angular diameter distance and the luminosity distance via the relation \( D_{\rm L}(z) = D_{\rm A}(z)(1+z)^2 \). Our analysis is based on the latest baryon acoustic oscillation (BAO) measurements provided by the Dark Energy Survey (DES), the Baryon Oscillation Spectroscopic Survey (BOSS)/Extended BOSS (eBOSS), and the Dark Energy Spectroscopic Instrument (DESI) surveys. Specifically, an unbiased test of the CDDR is performed through a novel, model-independent method inspired by the two-point diagnostic approach, with DES-SN5YR and Pantheon type Ia supernova (SN Ia) sample reconstructed using the Artificial Neural Network (ANN) technique. This methodology effectively eliminates all nuisance parameters, including the sound horizon scale \( r_{\rm d} \) from BAO and the absolute magnitude \( M_{\rm B} \) from SN Ia. A set of \( N-1 \) independent CDDR ratios \( \eta_{ij} \) are constructed for statistical analysis. At the current observational level, no significant deviation from the CDDR is observed at low redshifts, whereas we find positive evidence ($>2\sigma$ C.L.) of deviation from the CDDR at two high redshifts ($z=2.33$ and $z=2.334$). Therefore, our results confirm that the BAO measurement provides a powerful tool to test such fundamental relation in modern cosmology.

\end{abstract}


\section{Introduction}\label{sec:1}

As the most fundamental relation in modern cosmology, the cosmic distance duality relation (CDDR) is a fundamental principle essential for interpreting geometrical distance measurements in astronomical observations. Such Etherington’s reciprocity law directly links the luminosity distance $D_{\rm L}$ and the angular diameter distance $D_{\rm A}$ as $D_{\rm L}(z)=D_{\rm A}(z)(1+z)^2$ \citep{CDDR1933}, where $z$ is the redshift. This relationship holds true under the conditions where a metric theory of gravity describes spacetime, photons travel along unique null geodesics, and the number of photons is conserved \citep{CDDR2007}. Therefore, the possible deviation from the CDDR could indicate the existence of systematics in observations or even new physics, such as the coupling of photons with non-standard particles \citep{photon2004}, the variation of fundamental constants \citep{constant2013}, etc. Moreover, the CDDR has also been widely applied in measuring cosmic curvature using strong gravitational lensing systems \citep{SGL2017,SGL2019,SGL2020}, determining the geometry, gas mass density distribution and temperature distribution of galaxy clusters \citep{temperature2011,galaxy2011,gasmass2016}.

The methodology for testing the CDDR is conceptually straightforward: it involves measuring both the $D_{\rm A}(z)$ and the $D_{\rm L}(z)$ at the same redshift $\eta(z)=D_{\rm L}(z)/D_{\rm A}(z)(1+z)^2$. Several works have employed such an approach to test its validity. Type Ia supernovae (SN Ia) are regarded as standard candles that could provide the measurements of $D_{\rm L}(z)$, while $D_{\rm A}$ is typically inferred from strong gravitational lensing systems \citep{SGLSN2018}, galaxy clusters \citep{Holanda2010,2011Galaxy}, the angular size of ultra-compact radio sources \citep{2018radio}, and baryon acoustic oscillations (BAO) \citep{2015BAO,2020BAO}. Interestingly, \citet{2019GW} have explored the potential of deriving $D_{\rm L}$ measurements from gravitational wave sources, imposing more restrictive limits on possible deviations from CDDR. However, in order to conduct more robust tests of the CDDR, it is necessary to mitigate the effects and uncertainties associated with the nuisance parameters from the observed data, namely the \( D_{\rm L} \) and the \( D_{\rm A} \). In the past year, many valuable approaches to testing CDDR have been put forward. For instance, it has been suggested that nuisance parameters, which characterize SN Ia light curves, significantly contribute to the uncertainties in the final results \citep{Linuisance,weinuisance}. Angular distance measurements from ultra-compact radio sources suffer the same problem, i.e., the linear size of the compact structure in radio quasars need not be calibrated. Following this direction, many new methods for CDDR tests have also been developed in the literature \citep{Martinelli2020,2021Qin,Tang2023,BAOdata,Jesus2025,2025Huang}. Specially, the CDDR parameter $\eta(z)$ in conjunction with the approach introduced by \citet{tonghua2023}, which employs the ratio \( \eta(z_i)/\eta(z_j) \), allows for the elimination of the effects and uncertainties associated with these nuisance parameters. With the data release of Dark Energy Spectroscopic Instrument (DESI) survey, some studies proposed that the Hubble tension could be transformed into possible deviation from the CDDR by incorporating $\rm SH0ES$ and $\rm BBN$ calibration \citep{Teixeira2025}. A notable discrepancy has been revealed \citet{keil2025} between the results from model-independent Genetic Algorithm (GA) approaches and those derived from parametric methods, highlighting the critical role of model-independent analysis in cosmological studies.

In this paper, we aim to present new tests of the CDDR by utilizing $\eta(z_i)/\eta(z_j)$ as cosmological observables. The $D_{\rm L}$ measurements are inferred from the latest SN Ia samples (DES-SN5YR and Pantheon), while the $D_{\rm A}$ observations are obtained from 2D BAO dataset from the public data releases (DR) of the Sloan Digital Sky Survey (SDSS), BOSS, and DES Y6, as well as 3D BAO datasets created by combining the latest DESI Y1 and Y2 data, the DES Y6 and BOSS/eBOSS datasets. Specifically, we achieve an unbiased measurement of the CDDR through a novel model-independent methodology inspired by the two-point diagnostic. This approach could effectively eliminate all nuisance parameters about the sound horizon ($r_{\rm d}$) from BAO and the absolute magnitude ($M_{\rm B}$) from SN Ia. Since the redshift range of SN Ia do not entirely overlap with those of BAO, the Artificial Neural Network (ANN) algorithm is used to reconstruct the potential evolution of CDDR with redshift. It is important to note that such a data-driven approach is entirely cosmological-model-independent and does not assume any specific characteristics of the observed data. In the framework of our new methodology, an uncorrelated calculation method using $N-1$ independent ratios at different redshifts will be employed. Such a procedure would contribute to determining positive evidence of CDDR failure at specific redshifts.


\section{Data and Methodology}\label{Sec:2}


\subsection{Data}
\begin{figure}[!t]
    \centering
    \includegraphics[width=0.49\textwidth]{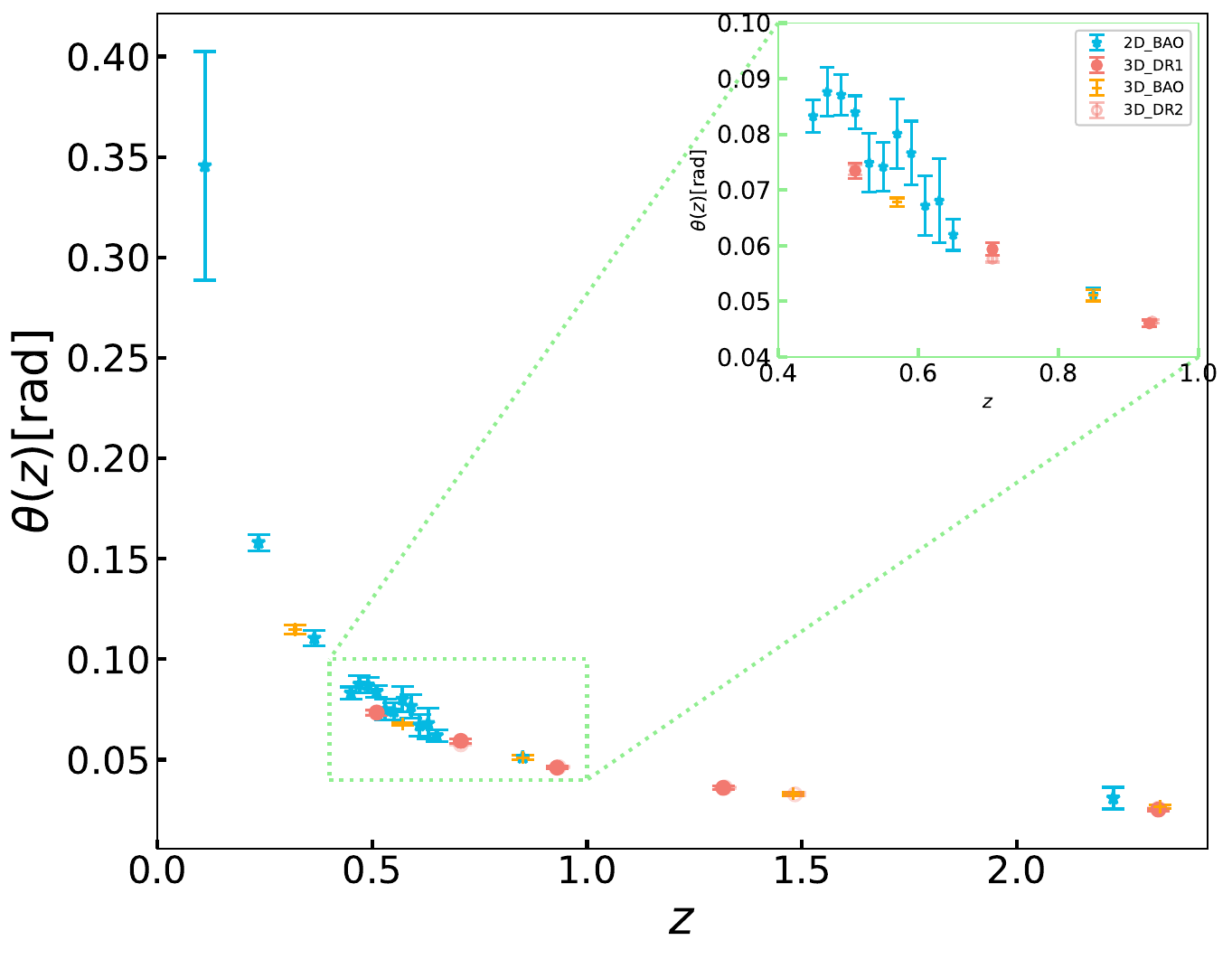}
    \caption{The complete BAO sample used in this analysis, following the relation of $\theta(z)~[\rm rad] = r_d / \left[D_{\rm A}(z)(1+z)\right]$. The blue and orange points represent the 2D BAO and 3D BOSS measurements, the solid red points correspond to 3D DESI DR1 data, and the hollow light-red points denote the 3D DESI DR2 data. The inner panel illustrates the data points and associated error bars within the redshift range of $0.4<z<1$.} 
    \label{fig:1}
\end{figure}

The BAO, which arise from interactions between baryons and photons in the early universe, serve as a standard ruler, observable through galaxy distributions in large-scale structures. By observing the distribution of galaxies and quasars at different redshifts, the distribution of matter in the universe can be measured, allowing constraints on cosmological parameters. The accuracy of this standard ruler is crucial for cosmological studies involving BAO. Galaxy surveys have been able to measure the angular BAO scale $\theta$. The relationship is given by
\begin{equation}
   D_{\rm A}(z)=\frac{r_{\rm d}}{\theta(z)(1+z)},
\label{eq:1}
\end{equation}
where $r_{\rm d}$ is the sound horizon at redshift $z$.
The 16 transversal BAO (2D\_BAO) measurements are obtained by extracting the BAO signal from the analysis of the angular two-point correlation function or the angular power spectrum, measuring the angular position of the BAO peak without requiring a fiducial cosmological model \citep{3DBAOsoure}. Additionally, five anisotropic BAO (3D\_BAO) measurements utilize the radial clustering signal imprinted on the large-scale structure of the universe. These measurements are comprehensively summarized in Tables 1 and 2 of \citet{BAOdata} and are employed in this paper. These values are obtained using public data from the SDSS namely DR7, DR10, DR11, DR12, DR16, DR12Q, DR16Q (quasars), and DES Y6, without assuming any fiducial cosmological model. Precisely because these BAO measurements are performed using cosmology-independent methods, their uncertainties are larger than those obtained using fiducial cosmologies.
In addition, the latest DESI Year 1 (3D\_DR1) and Year 2 (3D\_DR2) data releases are also utilized \citep{DESI2024,DESI2025}. These datasets incorporate measurements from Luminous Red Galaxies (LRG), Emission Line Galaxies (ELG), quasars (QSO), and Lyman-$\alpha$ forest (Ly$\alpha$) tracers across the redshift range $0.1 < z < 2.4$. Compared with 3D\_DR1, the 3D\_DR2 data show improved redshift confidence, more accurate measurements at the same redshifts, and new measurements at the QSO regime ($z=1.484$). The DESI data release \citep{DESI2024III} has significantly extended the redshift coverage beyond BOSS, with the number of objects observed in these releases far exceeding the BOSS sample size. These novel DESI BAO measurements have been extensively applied in determining the Hubble constant, cosmic curvature, and dynamics of dark energy \citep{Guo_2025,Lui_2025apjl,keil2025}. The BAO angular scale measurements used in this paper are presented in Fig.~\ref{fig:1}.


As a reliable and well-characterized cosmic probe, SN Ia span a redshift range that approximately coincides with BAO, thereby providing an independent measurement of $D_{\rm L}$ required for this analysis. More specifically, the luminosity distance at redshift $z$ can be derived as
\begin{equation}
   D_{\rm L}(z) = 10^{(m_{\rm B, c o r r}-{M_{\rm B}})/5-5} \, \text{Mpc},
   \label{eq:4}
\end{equation}  
where $m_{\rm B,corr}$ represents the corrected apparent magnitude, and $M_{\rm B}$ denotes the absolute magnitude in the $\rm B$-band. A distinct approach known as BEAMS with Bias Corrections (BBC) \citep{2017BBC} is adopted
\begin{equation}
    m_{\rm B, corr} = m_{\rm B} + \alpha^* \cdot X_1 - \beta \cdot C,
    \label{eq:3}
\end{equation}
where $m_{\rm B}$ refers to the measured peak magnitude in the B band rest frame, $x_1$ is the light curve stretch parameter, and $C$ denotes the light curve color parameter. $\alpha^*$ and $\beta$ are two nuisance distance estimate parameters. As part of their Year 5 data release, the Dark Energy Survey (DES) recently published results from a new, homogeneously selected sample of 1635 photometrically classified SN Ia with redshifts spanning \( 0.1 < z < 1.3 \) \citep{DES2024}. This sample is complemented by 194 low-redshift SN Ia (shared with the Pantheon-Plus sample) in the range \( 0.025 < z < 0.1 \). We refer to this dataset as DESY5. At high-redshift regime, the recent SN Ia sample called Pantheon is adopted \citep{Scolinic2018}, which encompasses observations from the PanSTARRS1 (PS1) Medium Deep Survey, the Sloan Digital Sky Survey (SDSS), the Supernova Legacy Survey (SNLS), and several low-redshift and Hubble Space Telescope samples. This comprehensive dataset includes 1048 SN Ia within a redshift range of $0.01 < z < 2.3$, providing a robust basis for investigating cosmological parameters and testing the validity of the CDDR. However, it is noteworthy that despite the rapid increase in the sample size of SN Ia (which has significantly improved the analysis and mitigation of systematic uncertainties), nuisance parameters such as the absolute magnitude ($M_{\rm B}$) still need to be optimized, potentially affecting the accuracy of the CDDR validation.



\begin{figure*}[!t]
  \centering
  \begin{minipage}[b]{0.49\textwidth}
    \includegraphics[width=0.95\textwidth]{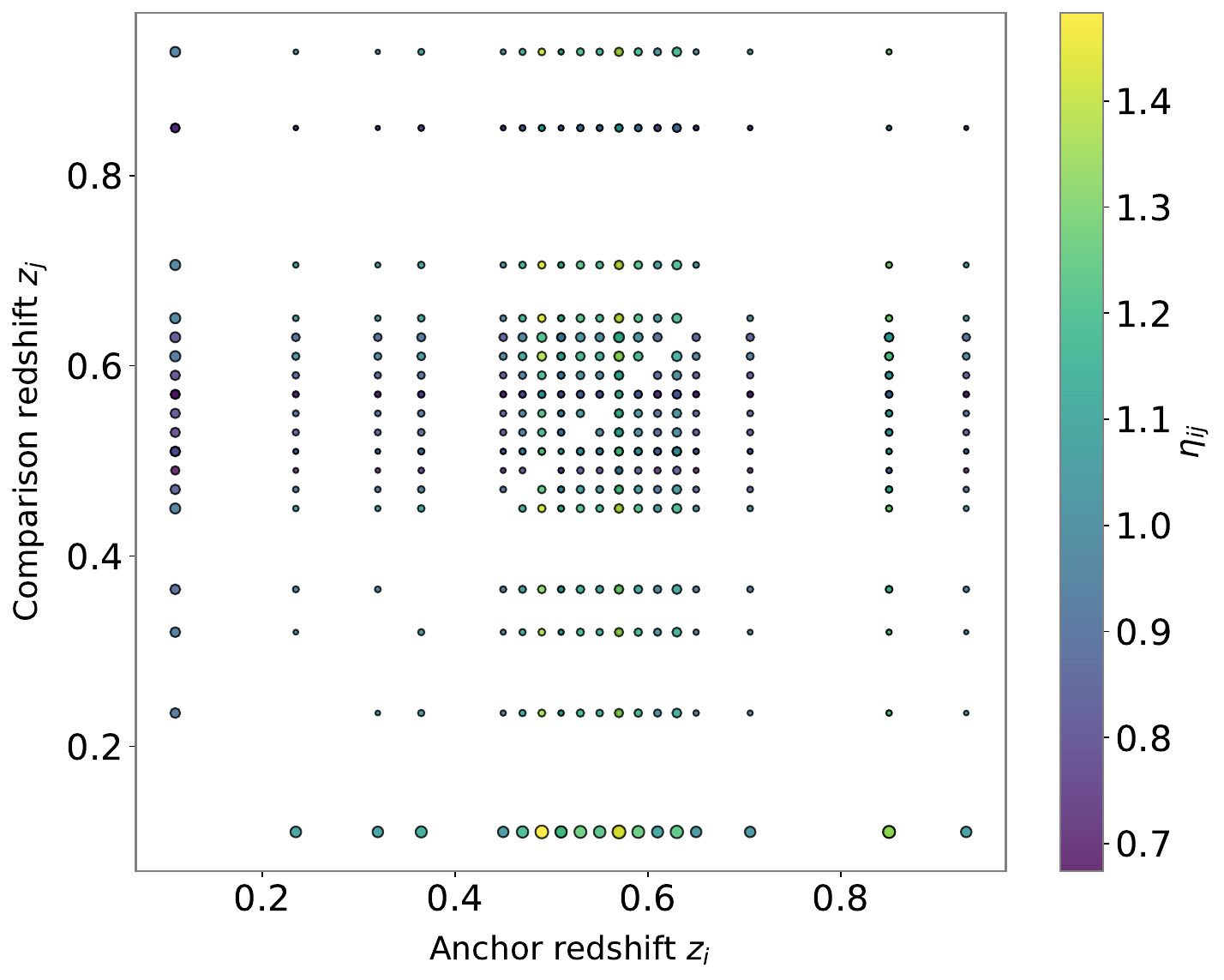}
  \end{minipage}
  \hfill 
  \begin{minipage}[b]{0.49\textwidth}
    \includegraphics[width=0.95\textwidth]{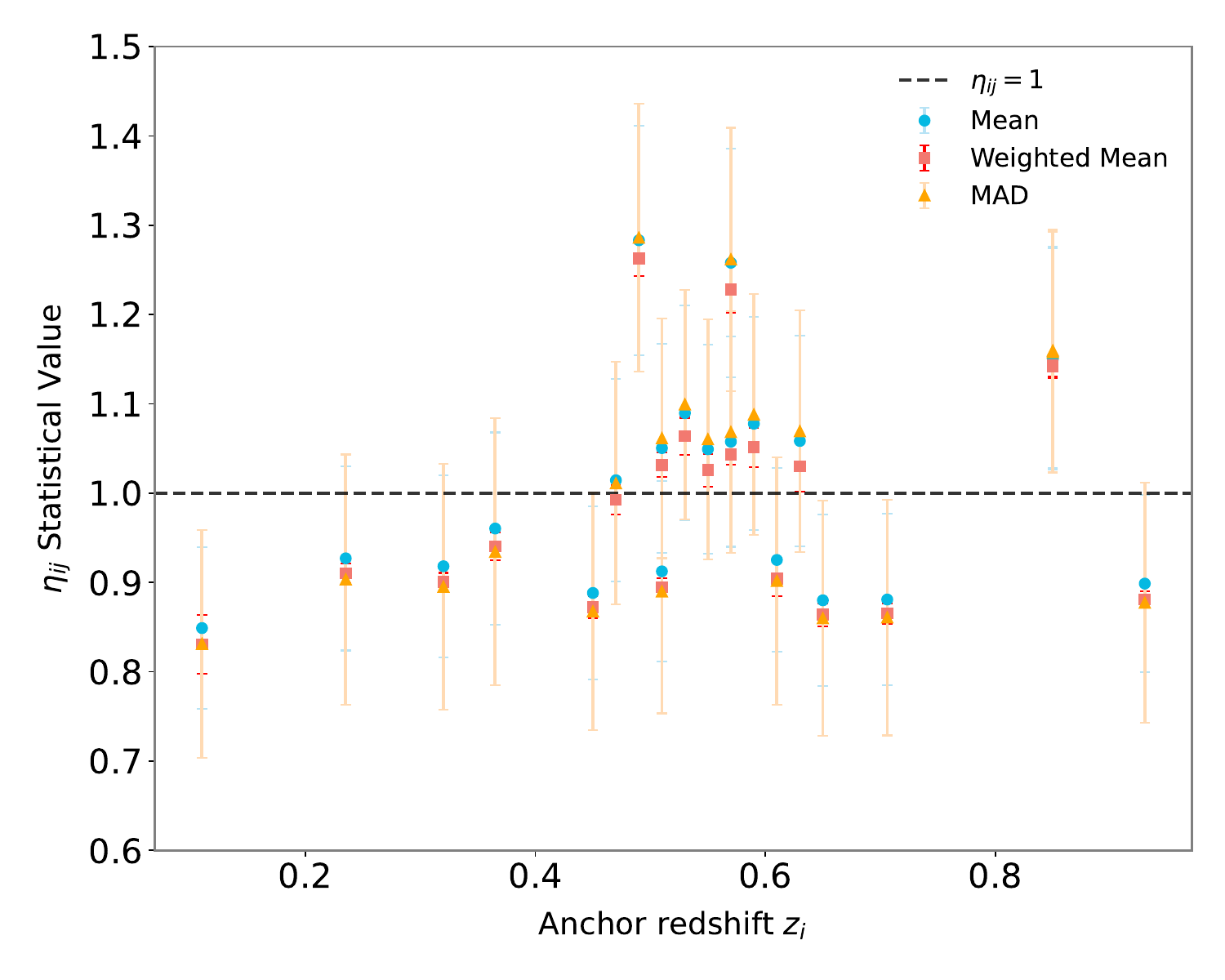}
  \end{minipage}
  \caption{A 2D bubble plot of the $\eta_{ij}$ two-point diagnostics derived from 21 data pairs between DESY5 and the combined BAO datasets (\texttt{3D\_BAO} and \texttt{3D\_DR1}). The color of each bubble indicates the value of $\eta_{ij}$, while the bubble size is proportional to the corresponding error bar (left panel). The results obtained from three different statistical methods (anchored at each of the 21 $z_i$), are also presented (right panel).}
  \label{fig:3}
\end{figure*}

\begin{figure*}[!t]
  \centering
  \begin{minipage}[b]{0.49\textwidth}
    \includegraphics[width=0.95\textwidth]{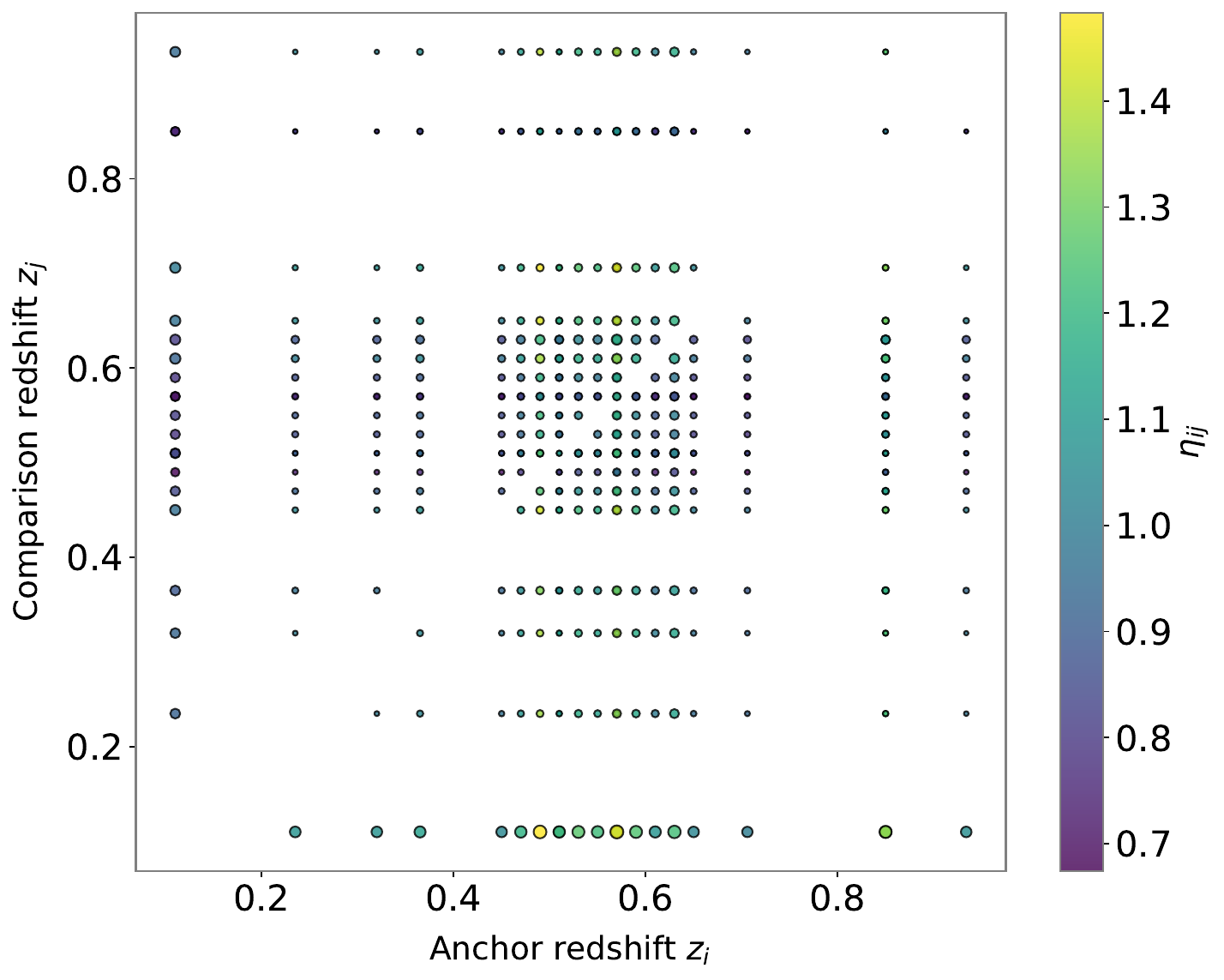}
  \end{minipage}
  \hfill 
  \begin{minipage}[b]{0.49\textwidth}
    \includegraphics[width=0.95\textwidth]{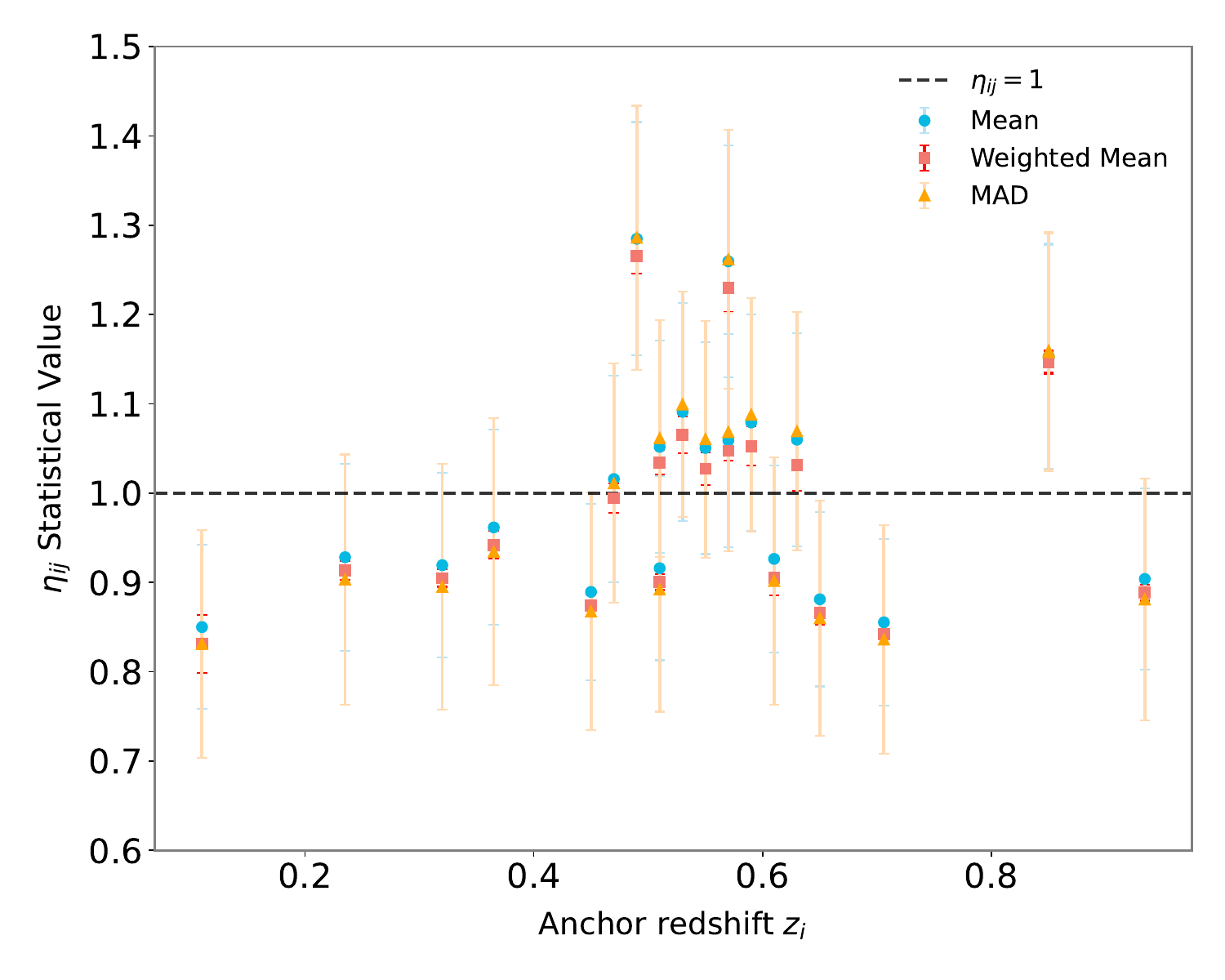}
  \end{minipage}
  \caption{The same as Figure 2, but for 21 data pairs between DESY5 and the combined BAO datasets (\texttt{3D\_BAO} and \texttt{3D\_DR2}).}
  \label{fig:4}
\end{figure*}

\begin{figure}[!t]
  \centering
  \includegraphics[width=0.47\textwidth]{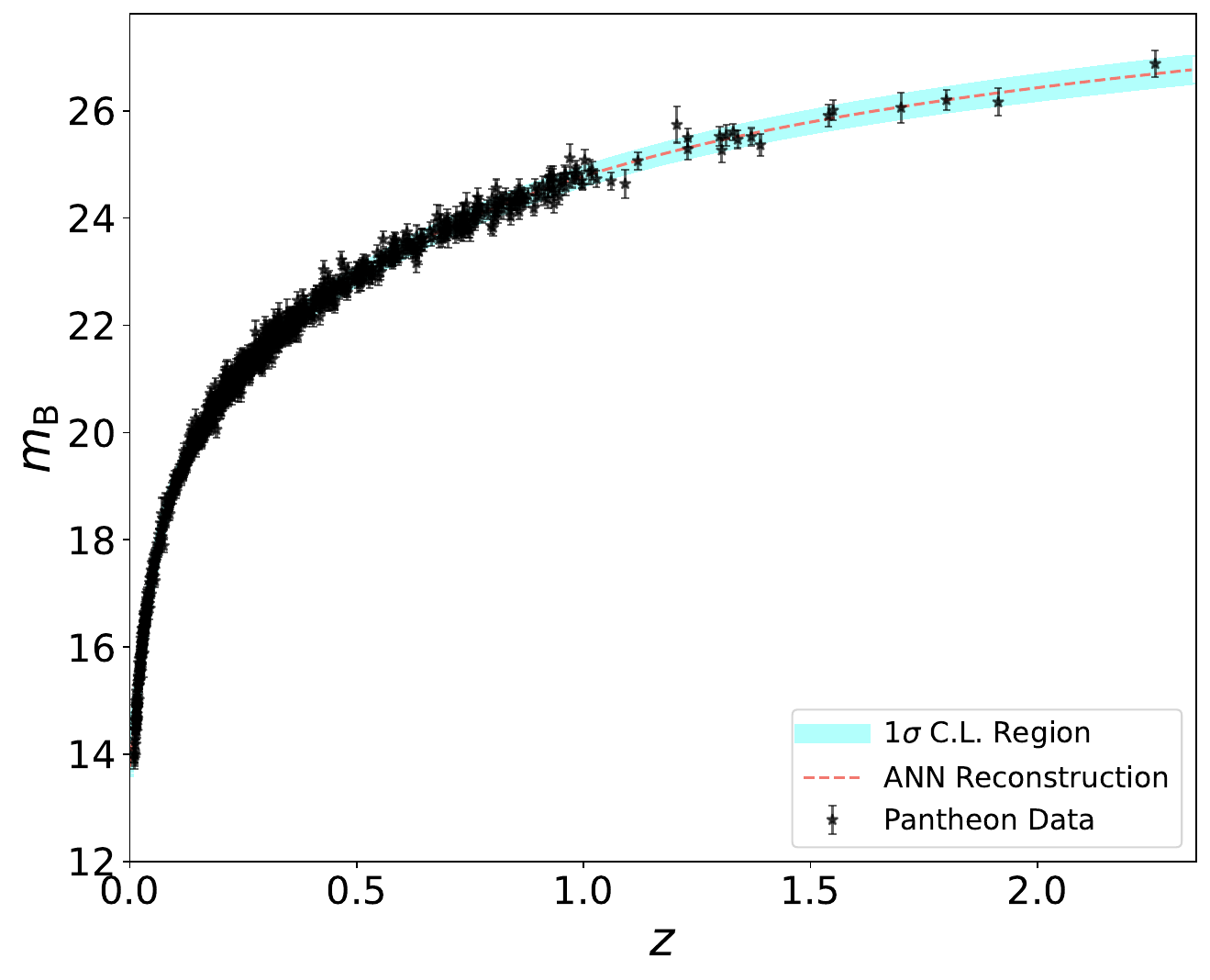}
  \caption{The Pantheon dataset and the reconstructed apparent magnitude ($m_{\rm B}$) using ANN. The blue shaded area represents the $1\sigma$ confidence region of the ANN reconstruction, while the black points with error bars show the Pantheon data points. The ANN reconstruction closely follows the trend of the observational data, providing a smooth and continuous representation of the apparent magnitude-redshift relation up to $z \approx 2.4$.}
  \label{fig:2}
\end{figure}

\subsection{Methodology}
To test the CDDR using the distance relation, a parameterized form is generally adopted, expressed as
\begin{equation}
    \eta(z) = \frac{D_{\rm L}(z)}{D_{\rm A}(z)(1+z)^2},
    \label{eq:5}
\end{equation}
where $D_{\rm A}$ is derived from the combination of transverse and anisotropic distances measured by BAO, without assuming any fiducial cosmological model. SN Ia observations provide a large number of relatively precise luminosity distance ($D_{\rm L}$) measurements. With the continuous expansion of supernova samples, their accuracy is also progressively improving. Consequently, an expression incorporating both can be readily formulated
\begin{equation}
    \eta(z)=\frac{\theta(z)10^{(m_{\rm B, \rm corr}-{M_{\rm B}} )/ 5-5}}{(1+z) r_{\rm d}},
    \label{eq:6}
\end{equation}
It is important to acknowledge that the measurement of the absolute magnitude ($M_{\rm B}$) and the sound horizon ($r_{\rm d}$) is influenced by various factors \citep{chen2024}, including the evolution of galaxy distribution and the B-band absolute magnitude of SN Ia, which depends on the mass of the progenitor stars. These factors have historically contributed to discrepancies in tests of the CDDR. To mitigate the impact of these two nuisance parameters, a novel method, $\eta(z_i)/\eta(z_j)$, was proposed by 
\begin{equation}
    \eta_{ij}=\eta\left(z_{i}\right) /\eta\left(z_{j}\right)=\frac{\theta\left(z_{i}\right)\left(1+z_{j}\right)}{\theta\left(z_{j}\right)\left(1+z_{i}\right)} 10^{\Delta m/5},
    \label{eq:7}
\end{equation}
where $\Delta m=m_{\rm B, \rm corr}(z_i)-m_{\rm B, \rm corr}(z_j)$, and $i$ and $j$ denote the indices of BAO and SN Ia, respectively \citep{tonghua2023}. The formula highlights an innovative method for examining the CDDR, which is exclusively based on observational data, thereby eliminating reliance on prior assumptions regarding the absolute magnitude ($M_{\rm B}$) or the sound horizon ($r_{\rm d}$). This method is rooted in the two-point diagnostic strategy, which has been widely employed to assess discrepancies between the $\Lambda{\rm CDM}$ model and various alternative dark energy models, including those incorporating dynamic dark energy components \citep{twopointzheng,2024liutwo}. As such, the coefficient $\eta_{ij}$ in the formula is independent of nuisance parameters, ensuring that the results remain unaffected by extraneous uncertainties. The fundamental advantage of this method lies in its ability to circumvent nuisance parameters, thereby systematically eliminating potential sources of ambiguity. Identifying redshifts that mutually overlap and enhance each other’s utility constitutes a crucial aspect of the analysis. To achieve this, two techniques are employed in which nearby supernovae are utilized to impose stringent constraints. The first strategy, which utilizes nearby SN Ia as reference points, imposes a strict redshift matching criterion of \( \Delta z < 0.005 \). Such a criterion is essential for achieving precise alignment between the redshifts detected in BAO and those of DESY5 SN Ia \citep{Holanda2010}. However, the restricted redshift coverage of DESY5 makes it difficult to match  higher-redshift BAO data. To overcome this limitation, the Pantheon dataset spanning a significantly wider redshift range will be used, combined with advanced machine learning algorithms for data reconstruction.

Machine learning algorithms have been widely used in cosmological research. For instance, convolutional neural networks have proven effective in extracting information from weak lensing mass maps, offering an alternative to traditional power spectrum analyses \citep{deepleaning1}. Furthermore, these networks have been utilized to forecast multistage solar flares within a 24-hour timeframe \citep{deepleaing2}, and to facilitate the discovery of strong gravitational lenses \citep{deeple4, deeple5, deeple3}. The aim is to ensure that the redshifts of the reconstructed dataset precisely match those of the BAO data. The publicly available code, Reconstruct Functions with ANN, referred to as ReFANN (\url{https://github.com/Guo-Jian-Wang/refann}) \citep{ANNwang}, which has demonstrated strong potential in resolving cosmological challenges and constraining key cosmological parameters, is utilized in this paper to reconstruct the Pantheon dataset \citep{fluriddr,xuddr}. ANN computational models, inspired by the human brain's information-processing mechanisms, consist of an input layer, one or multiple hidden layers, and an output layer. During the training phase, the ANN adjusts its weights and biases by evaluating the discrepancy between the predicted output $\bm{f_w}$, the bias term $\bm{b(x)}$, and the actual target value $\bm{y}$. This discrepancy is measured by a loss function $\bm{L}$, which quantifies the deviation between the predicted and actual values. The objective is to minimize this loss function by iteratively updating the weight matrix $\bm{W}$ and the bias vector $\bm{b}$, thereby reducing the gap between the network's predictions and the true outcomes \citep{ANNguocheng1}. This process typically involves gradient descent or its variants, which adjust the weights based on the gradient of the loss function with respect to the weights.

Through this mechanism, the ANN iteratively refines its parameters to minimize the error, leading to increasingly accurate predictions that closely align with the target values. As a data-driven technique, an ANN does not assume any specific distribution of the data, such as Gaussianity. Consequently, with a well-designed network architecture, an ANN can effectively approximate the underlying distribution of the input data \citep{ANNwang}.
We also evaluate the $\chi^2$ values for the supernova observational data and the ANN-reconstructed data (at the same redshift) against the $\Lambda$CDM model, yielding values of $\chi^2=1057$ and $\chi^2=850$, respectively. Therefore, one could reach a careful conclusion that the ANN reconstruction shows no indication of overfitting. Combining the BAO data together with the SN Ia sample (based on the redshift-match and ANN methods), one could directly obtain the ratio of CDDR parameter $\eta_{ij}$. The uncertainty of $\eta_{ij}$ is derived from the standard uncertainty propagation formula, which accounts for contributions from multiple independent observed variables. These sources encompass uncertainties in the corrected apparent magnitudes, $\sigma_{m_{\rm B, corr}}$, as well as uncertainties in the observed angular diameter distance and in angular size measurements $\sigma_\theta$.

\section{Results and Discussion}\label{Sec:3}
For each BAO data point included in the analysis, nearby SN Ia with $\Delta z < 0.005$ are selected to minimize systematic errors due to redshift mismatches. Two-point diagnostics involving the CDDR parameters, \( \eta(z_i) / \eta(z_j) \), are introduced as a model-independent tool for testing the CDDR. For each redshift, \( N-1 \) independent ratios are constructed by anchoring at that redshift and comparing it with all others, enabling a robust statistical evaluation. This method relies solely on observational data and does not assume any cosmological model. To implement this approach, the $\eta_i$ values at different redshifts are systematically used as anchors to generate $N$ groups, each containing $N-1$ independent ratio pairs, which are then subjected to statistical analysis. Moreover, it is crucial to recognize that the validity of the CDDR inherently implies that the ratio of its associated parameters should be invariantly equal to unity ($\eta_{ij}=1$). While the observed ratio $\eta_{ij}$ may indeed be unity, this does not automatically imply that the intrinsic value of $\eta$ is also unity. However, if $\eta_{ij} \neq 1$, it will provide important evidence indicating a potential violation of the CDDR. 

To better illustrate the significance of our results, three different statistical methods are implemented for comparative analysis. The standard deviation is first applied to compute the results, which is widely recognized as a traditional metric for quantifying the dispersion of measurements. However, one should note that the standard deviation can be inflated by the presence of outliers or extreme values, which may lead to a misinterpretation of the data results. With standard deviation, which uses the square of the distance from the data to the mean, larger deviations are weighted more heavily, and the impact of outliers on the results cannot be ignored. In certain instances, the significance of data points is not uniform, necessitating a weighting system that assigns relative importance to each point, either based on theoretical insights or its overall significance within the dataset. To improve the statistical robustness of the analysis and ensure that each data point's contribution is adequately accounted for, the weighted mean is initially employed for evaluation \citep{2003dataerror}. The Shapiro–Wilk test is performed on the datasets of \( \eta_{ij} \) obtained from redshift matching and ANN reconstruction, revealing the presence of datasets with \( p \leq 0.05 \). This indicates the the presence of possible deviation from Gaussian distribution in the $\eta_{ij}$ dataset. Therefore, the median absolute deviation (MAD) is introduced, which is particularly effective in handling asymmetric distributions and mitigating the impact of extreme values on statistical analysis \citep{peterMAD2012}. By the very definition of the median, it is inherently designed such that the chance of any single measurement out of $N$ independent observations exceeding the median value is exactly $50\%$. As a result, the likelihood of any given observation, being the n-th out of a total $N$, exceeding the median can be described by a binomial distribution. This is mathematically represented as $P(N > n) = \binom{N}{n} \left(\frac{1}{2}\right)^N$, where \( \binom{N}{n} \) is the binomial coefficient, reflecting the number of ways to choose $n$ successes in $N$ trials. This statistical property facilitates the computation of the $68\%$ confidence regions surrounding the median value as a robust statistic. More specifically, the calculation of MAD involves determining the differences between each data point and the median of the sample, and thus enables comparison with metrics such as the standard deviation. In this analysis, we turn to the normalized MAD which is expressed as \citep{MAD2024}
\begin{equation}
   \text{MAD}=1.48 \times \operatorname{median}\left(\left|\eta^n_{ij}-\eta_{ij}^{\rm m}\right|\right),
   \label{eq:10}
\end{equation}
where $\eta^n_{ij}$ represents the n-th value and $\eta^{m}_{ij}$ denotes the median value of the $\eta^n_{ij}$ dataset.

\begin{figure*}[!t]
  \centering
  \begin{minipage}[b]{0.49\textwidth}
    \includegraphics[width=0.95\textwidth]{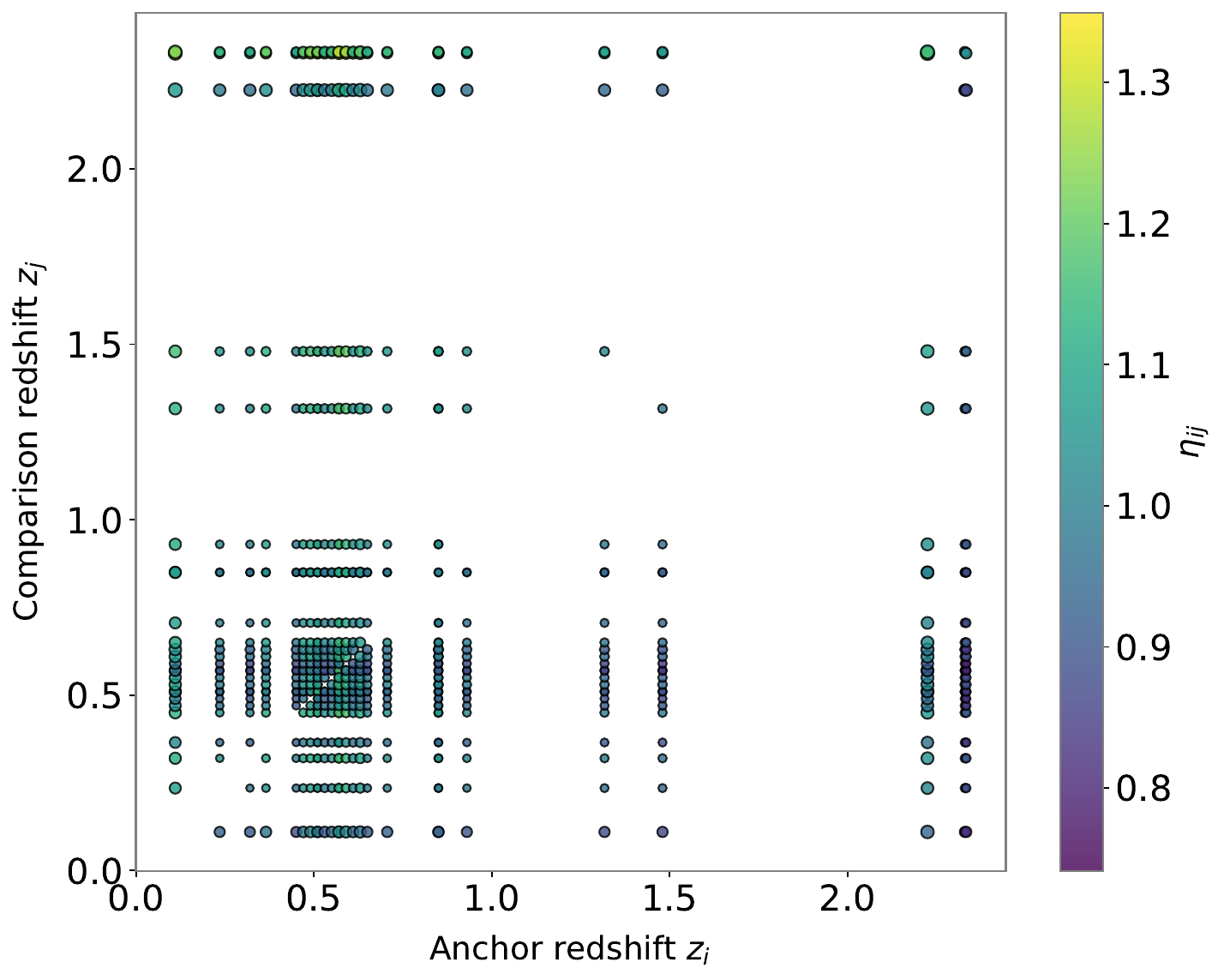}
  \end{minipage}
  \hfill 
  \begin{minipage}[b]{0.49\textwidth}
    \includegraphics[width=0.95\textwidth]{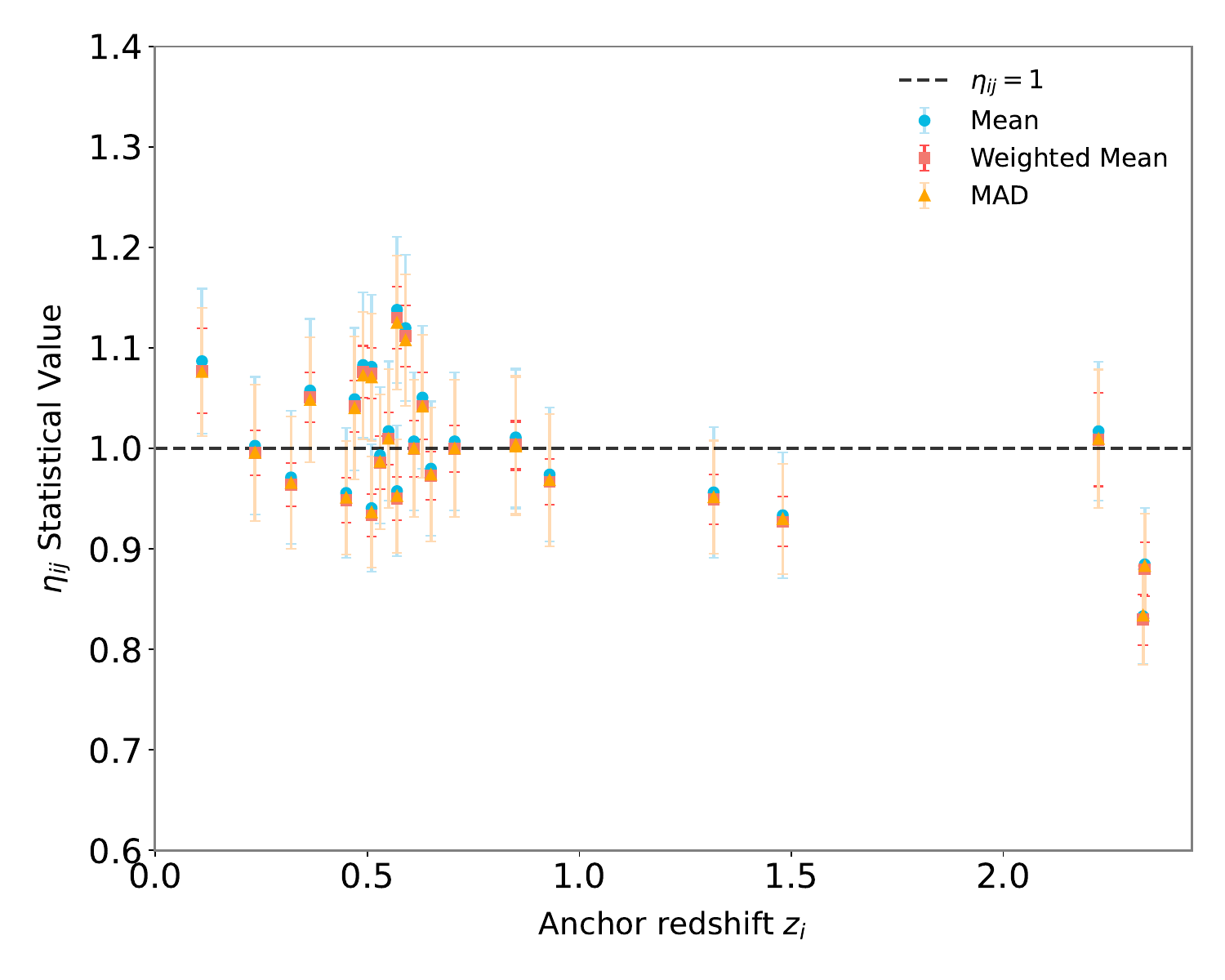}
  \end{minipage}
  \caption{A 2D bubble plot of the $\eta_{ij}$ two-point diagnostics derived from 26 data pairs between ANN-reconstructed Pantheon sample and the combined BAO datasets (\texttt{3D\_BAO} and \texttt{3D\_DR1}). The color of each bubble indicates the value of $\eta_{ij}$, while the bubble size is proportional to the corresponding error bar (left panel). The results obtained from three different statistical methods (anchored at each of the 26 $z_i$), are also presented (right panel).}
  \label{fig:5}
\end{figure*}

\begin{figure*}[!t]
  \centering
  \begin{minipage}[b]{0.49\textwidth}
    \includegraphics[width=0.95\textwidth]{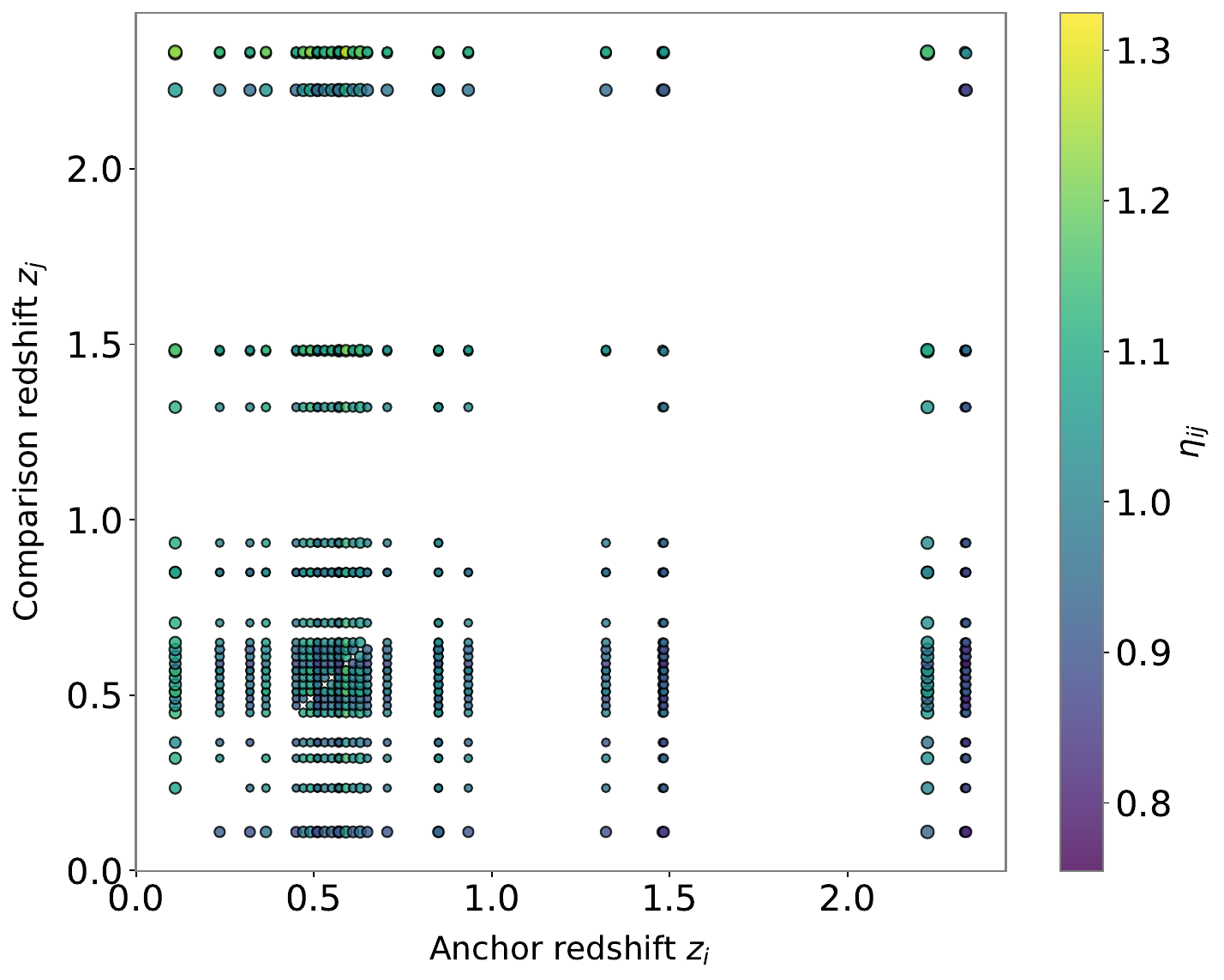}
  \end{minipage}
  \hfill 
  \begin{minipage}[b]{0.49\textwidth}
    \includegraphics[width=0.95\textwidth]{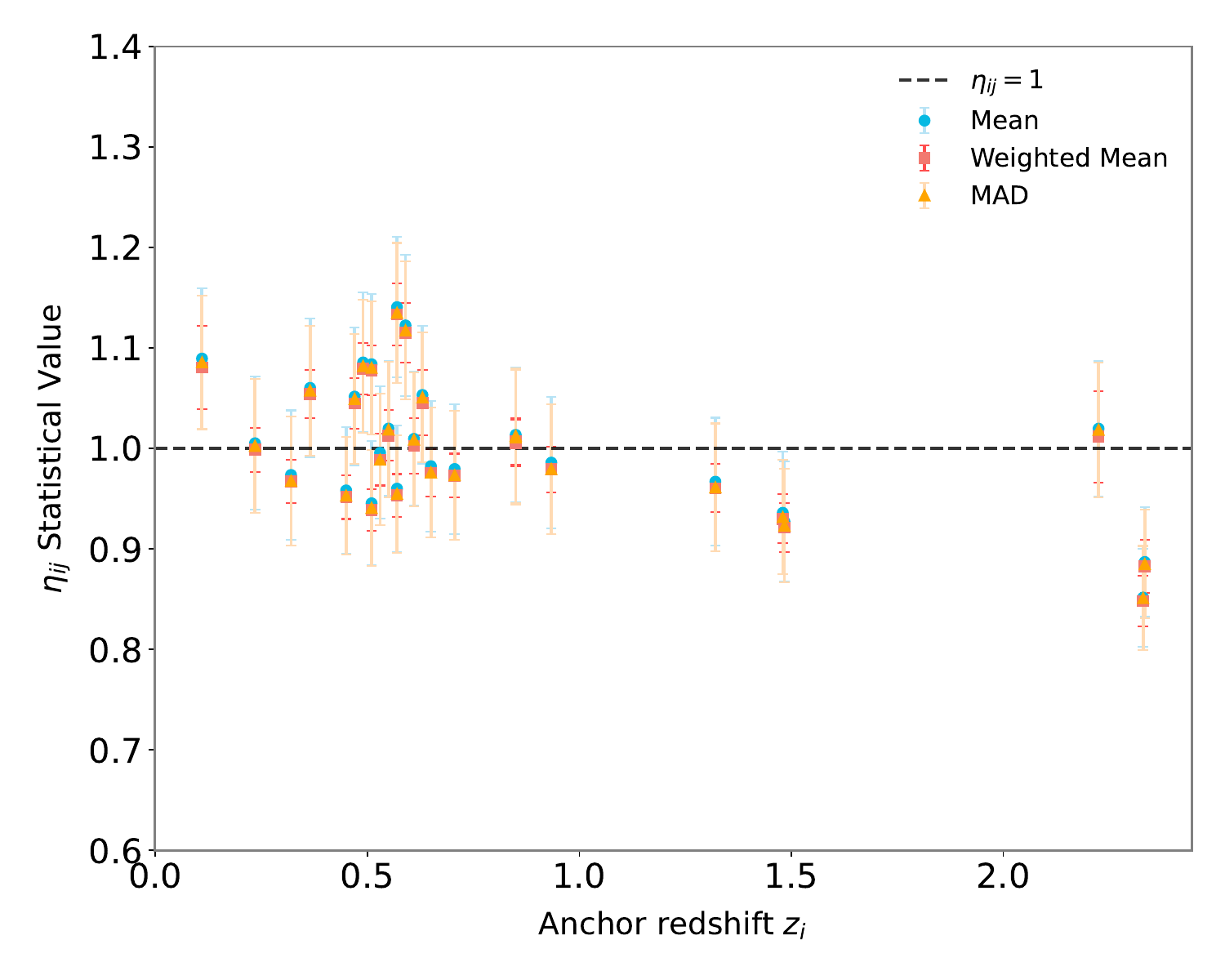}
  \end{minipage}
  \caption{The same as Figure 5, but for 27 data pairs between ANN-reconstructed Pantheon sample and the combined BAO datasets (\texttt{3D\_BAO} and \texttt{3D\_DR2}).}
  \label{fig:6}
\end{figure*}

Our analysis begins with the most recent DESY5 dataset, in combination with with BAO measurements from 3D\_BAO and 3D\_DR1. We present these diagnostics together with their uncertainties in Fig.~\ref{fig:3}. The left panel displays 21 $\eta_{ij}$ ratios constructed from various redshift combinations ($z_i$, $z_j$), while the right panel presents the deviation statistics of these ratios (anchored at $z_i$), under three distinct statistical approaches. A comparative analysis incorporating 3D\_BAO and 3D\_DR2 data is also provided in Fig.~\ref{fig:4}. It is observed that, at low redshifts, the analysis reveals that the vast majority of CDDR parameters remain consistent with the expected value, showing no significant deviations within the $2\sigma$ range. Although the weighted mean suggests possible signs of deviation from the CDDR at several redshifts, both the mean and the MAD values at those redshifts do not support such deviations. Such discrepancy may be attributed to the fact that the weighted mean is influenced by a small number of data points with large weights, particularly with the presence of extreme values. Notably, the 3D\_DR1 subsample includes the LRG1 and LRG2 points, which were previously reported by the DESI Collaboration as potentially deviating from standard cosmological model \citep{DESI2024}. However, in this analysis, these BAOs yield mean values and standard deviation of the two-point diagnostics of the CDDR parameter \( \eta_{ij}=0.916 \pm 0.103 \) and \( \eta_{ij}=0.855 \pm 0.093 \), respectively. The weighted mean and corresponding uncertainty are determined to be
\( \eta_{ij}=0.901 \pm 0.009 \) and \( \eta_{ij}=0.842 \pm 0.010\).  In the framework of MAD, the absolute median value and the median absolute deviation are calculated as $\eta_{ij}=0.892 \pm 0.137$ and $\eta_{ij}=0.836 \pm 0.128$. None of these results indicate statistically significant departures from the CDDR. An identical analysis is subsequently performed using the updated 3D\_DR2 data, with results showing no substantial difference. These findings are in perfect agreement with earlier studies \citep{etabierende2020,etabierende2021,etabierende2021b}, further reinforcing the consistency of the CDDR at low redshift across multiple statistical treatments and data releases.

Given the limited redshift coverage of the DESY5 dataset, we also turn to the Pantheon dataset that significantly expands in both sample size and redshift coverage. Fig.~\ref{fig:2} displays the ANN-reconstructed apparent magnitude ($m_B$) of the Pantheon sample, which provides the necessary ingredients at the redshifts corresponding to high-redshift DESI measurements. The corresponding results for 3D\_DR1, after incorporating ANN-reconstructed supernova data are presented in Fig.~\ref{fig:5}, from which a significant increase in the number of data points (26 $\eta_{ij}$ ratios) is observed. Our findings indicate that, for the 3D\_DR1 and Pantheon datasets, the vast majority of CDDR parameters remain consistent with the CDDR, without any statistically significant deviation. However, at two higher redshifts ($z=2.33$ and $z=2.334$), all three statistical methods demonstrate strong evidence of deviation ($\eta_{ij} \neq 1$) at $>3\sigma$ and $>2\sigma$, respectively. Considering that these data points have the highest weight in their respective combinations, it may indicate a potential failure of CDDR at these two redshifts. Further analysis is conducted on the updated 3D\_DR2 dataset, with results presented in Fig.~\ref{fig:6}. The newly added QSO data point at $z = 1.484$ shows a mean value and its corresponding uncertainty of $0.927 \pm 0.060$, a weighted mean of $0.921 \pm 0.024$, and the MAD of $0.923 \pm 0.056$. For the two high-redshift data ($z=2.33$ and $z=2.334$), positive evidence of deviation from the DDR ($\eta_{ij} \neq 1$) still exists ($>2\sigma$). This is the most unambiguous result of the current BAO datasets. We emphasize that the above analysis involves $N$ distinct statistical evaluations, which naturally introduces the well-known multiple comparison problem \citep{Benjamini95}. This effect can lead to an overestimation of the statistical significance associated with individual deviations, which are observed within the ensemble even in the absence of any true underlying bias. Although formal multiplicity corrections are not applied in this analysis \citep{10.1111/1467-9868.00346,Bretz2010MultipleCU}, we acknowledge that their implementation would likely reduce the nominal significance levels of the observed high-redshift deviations. Consequently, we interpret these particular findings with caution and emphasize that their current significance levels should be viewed within a broader statistical landscape. Future high-redshift BAO observations will be valuable in further assessing the stability and interpretation of these deviations.

In order to investigate the origin of such CDDR deviation, a sensitivity test is performed if the CDDR holds at redshifts $z=2.33$ and $z=2.334$. Fixing the absolute magnitude of SN Ia at $M_{\rm B}=-19.25$, one could derive the sound horizon of BAO as $r_{\rm d}=122.52 \pm 14.67$Mpc and $r_{\rm d} = 127.40 \pm 15.62$Mpc, respectively. There would be a $1.7\sigma$ and $1.3\sigma$ tension between these results and the best-fit value from \citet{plank2018}. When the sound horizon of BAO is fixed at $r_{\rm d}=147.09$Mpc, the absolute magnitude of SN Ia will change to $M_{\rm B}=-19.65 \pm 0.26$ and $M_{\rm B}=-19.56 \pm 0.27$. There would be a $1.6\sigma$ and $1.2\sigma$ tension between these results and the best-fit value from \citet{Riess2022}. One should note that, the final Ly$\alpha$ BAO measurement using the complete BOSS data sample (i.e., the cross-correlation of quasars with the Ly$\alpha$-forest flux transmission) at an effective redshift $z=2.33$, also exhibits a $2.3\sigma$ tension with the prediction of the best-fit $\Lambda$CDM cosmology from the Planck satellite \citep{2017A&A...608A.130D}. Such tension has reduced to $1.5\sigma$ with the combination of all data from BOSS and eBOSS \citep{duMas2020}. On the other hand, the DESI collaboration results indicated a higher central value of \(\Omega_m\) in the Pantheon+ dataset compared to the results obtained from DESI alone. This difference reaches statistical significance at the \(1.7\sigma\) level \citep{DESI2025}. This discrepancy indicates a notable inconsistency when these datasets are interpreted within the flat \(\Lambda\)CDM model, suggesting potential systematic or model-dependent factors that could have influenced these findings. In addition, observational systematic errors, such as imaging systematics, incomplete fiber assignment, redshift space distortions, nonlinear effects, and noise, may all impact redshift measurements and cluster abundance measurements. Moreover, systematic biases may arise during data processing and model fitting due to choices in target identification, weight assignment, and fitting methodologies, potentially affecting the reliability and precision of DESI measurements \citep{DESI2024}. Summarizing, our results confirm that the BAO measurement provides a powerful astrophysical tool, which has profound implications for the understanding of fundamental physics in modern cosmology.

\section*{Acknowledgments}

This work is supported by Beijing Natural Science Foundation No. 1242021; the National Natural Science Foundation of China under Grants Nos. 12021003, 12433001, and 12041301; the Fundamental Research Funds for the Central Universities. T. Liu was supported by National Natural Science Foundation of China under Grant No. 12203009; Chutian Scholars Program in Hubei Province (X2023007); Hubei Province Foreign Expert Project (2023DJC040).

\bibliography{main.bib} 

\end{document}